\documentclass[prb,twocolumn,floatfix,citeautoscript,nofootinbib,superscriptaddress]{revtex4}
\usepackage{amsbsy}
\usepackage{latexsym,epsfig,graphicx}
\usepackage{dcolumn}
\usepackage{graphicx}
\usepackage{subfigure}
\usepackage{comment}
\usepackage{color}
\usepackage{bm}
\usepackage{mathrsfs}
\usepackage{amsfonts}
\usepackage{amsmath}
\usepackage{color}
\usepackage{amssymb}
\usepackage{xspace}
\usepackage{epstopdf}
\usepackage{tabularx}
\usepackage{longtable}
\usepackage[colorlinks=true, letterpaper=true, pdfstartview=FitV, linkcolor=blue, citecolor=blue, urlcolor=blue]{hyperref}
\usepackage[normalem]{ulem}

\begin{document}

\title{Temperature-Induced Magnonic Chern Insulator in Collinear Antiferromagnets}
\author{Yun-Mei Li}
\email[Corresponding author: ]{yunmeili@xmu.edu.cn}
\affiliation{Department of Physics, School of Physical Science and Technology, Xiamen University, Xiamen 361005, China}
\author{Xi-Wang Luo}
\affiliation{CAS Key Laboratory of Quantum Information, University of Science and Technology of China, Hefei 230026, China}
\author{Kai Chang}
\affiliation{SKLSM, Institute of Semiconductors, Chinese Academy of Sciences, P.O. Box 912, Beijing 100083, China}

\begin{abstract}
  Thermal fluctuation in magnets will bring temperature-dependent self-energy corrections to the magnons,
  however, their effects on the topological orders of magnons is not well explored.
  Here we demonstrate that such corrections can induce a Chern insulating phase
  in two-dimensional collinear antiferromagnets with sublattice asymmetries by increasing temperature.
  We present the phase diagram of the system and show that the trivial magnon bands at zero temperature exhibit Chern insulating phase above
  a critical temperature before the paramagnetic phase transition.
  The self-energy corrections close and reopen the bandgap at $\Gamma$ or $\mathbf{K}$ points,
  accompanied by a magnon chirality switch and nontrivial Berry curvature transition.
  The thermal Hall effect of magnons or detecting the magnon polarization can give experimentally prominent signatures of topological transitions.
  We include the numerical results based on van der Waals magnet MnPS$_{3}$,
  calling for experimental implementation.
  Our work presents a new paradigm for constructing topological phase that is beyond
  the linear spin wave theory.
\end{abstract}

\maketitle

\section{Introduction}

The last twenty years have witnessed the extraordinary developments
of topological insulators and semimetals in the field of condensed matter physics~\cite{MZHasan,XLQi,NPArmitage,BABernevig,MSMiao,DZhang,SYXu,PDziawa,FZhao,IGarate,GAntonius,TImai,PAPuente,BMWojek}.
In analog to electronic systems, the topological phases have also been extended to the bosonic systems, such as
the photonic~\cite{LLu} and acoustic systems~\cite{JLu}. Magnons, quantized spin excitations in magnets, another
boson, are also proposed to host nontrivial topological phases,
realized in magnets with artificially designed structures~\cite{RShindou,YMLi1,ZHu2}, special crystal symmetries~\cite{LZhang,AMook1,SKKim,THirosawa,HKondo,YMLi2,XSWang,YSLu,RChisnell},
or the quantum fluctuations~\cite{AMook3}.
The emergence of edge or surface states immune to disorder and back scattering has great potential applications
for designing magnonic devices~\cite{AVChumak,VBaltz} with low dissipation and power consumption.

One of the key features in magnetic systems is the presence of magnon-magnon interactions (MMIs) and thermal fluctuation.
Their interplay would give rise to temperature-dependent nonlinear self-energy corrections to the magnons ~\cite{FJDyson,SHLiu,BGLiu,SSPershoguba,ZLi2,BWei,VVMkhitaryan,TOguchi}.
Recent two works~\cite{AMook3,YSLu} stated that the nonlinear corrections can drive a topological phase
transition of Dirac magnons hosting opposite Chern numbers at a critical temperature.
The other works discussed the magnon topology within the linear spin wave theory ~\cite{LZhang,AMook1,SKKim,THirosawa,HKondo,RShindou,YMLi1,YMLi2,XSWang,ZHu2,RChisnell,AMook2}
or only the magnon renormalization effect~\cite{BGLiu,SSPershoguba,ZLi2,BWei,VVMkhitaryan}.
None of them addressed the possibilities of realizing topological phases of magnons above a finite temperature $T_{c}$
while the bands below $T_{c}$ are topologically trivial with $T_{c}$ below the Curie or N\'{e}el temperature.
This is quite reasonable because the specific schemes responsible
for the topological phases at zero temperature are always
present or absent in the self-energies at finite temperature.
This picture explains why few works made explorations
on the construction of a topological phase at finite temperature.

In this paper, we show that increasing temperature can actually induce topological phase for magnons by considering
the two-dimensional collinear antiferromagnet MnPS$_{3}$ as the example.
We introduce sublattice asymmetric magnetic interactions induced in heterostructures,
breaking the $\mathcal{PT}$ symmetry and magnon band degeneracy. We find that at zero temperature,
the Chern insulating phase emerges, but only exists in a finite interval of the single-ion easy-axis anisotropy strength.
The self-energy corrections do not destroy this topological phase.
Outside the interval, the magnon bands are topologically trivial at $T=0$.
But as temperature increases, due to the self-energies,
the bandgap at $\Gamma$ or $\mathbf{K}$ points will be closed and reopen above a critical temperature $T_{c}$,
which is well below the N\'{e}el temperature.
The topological invariant, i.e., the Chern integer
of the acoustic branch changes from $0$ to $1$ across $T_{c}$.
We also find that the bandgap closing and reopening are accompanied by a magnon chirality switch
and nontrivial Berry curvature transition near $\Gamma$ or $\mathbf{K}$ points.
The thermal Hall effect gives prominent signatures for the topological phase transitions near $\Gamma$  point.
Also detecting the magnon polarization provides another experimental proofs.
Our proposal and conclusion are quite universal in collinear antiferromagnets, and can be extended to ferrimagnets.

This paper is organized as follows. In Sec. II, we present the model and use the finite-temperature field theory to deal with
the MMIs.  In Sec III, we calculate the magnon bands at both zero and finite temperatures and the corresponding topological invariant.
We present the phase diagram and discuss topological phase induced by increasing the temperature.
We also give the thermal Hall effect and magnon chirality switch during the topological phase transitions, which
can be probed in realistic experiments.
Finally, we summarize in Sec. IV.

\section{Model and methodology}

\begin{figure}[t]
  \centering
  \includegraphics[width=0.4\textwidth]{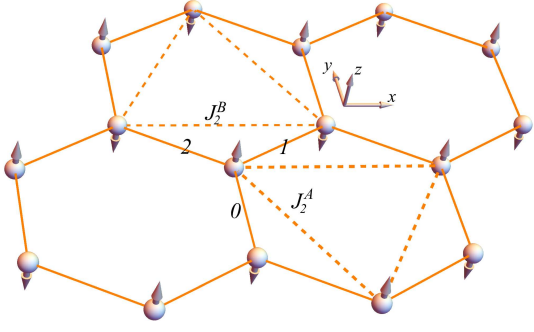}\\
  \caption{Illustration of the honeycomb antiferromagnet with sublattice asymmetry.
  The numbers denote the bond index. Spins on A-sublattice point up while on B-sublattice point down.
  The intra-sublattice second neighbor interactions on two sublattices are denoted as
  $J_{2}^{A}$ and $J_{2}^{B}$, respectively.}\label{fig1}
\end{figure}

We consider a honeycomb collinear antiferromagnet, as illustrated in Fig.~\ref{fig1}.
the spin interaction Hamiltonian is given by
\begin{eqnarray}\label{eq1}
  H&=&J_{1}\sum_{\langle ij\rangle}\mathbf{S}_{i}\cdot\mathbf{S}_{j}+I_{1}\sum_{\langle ij\rangle}S_{i}^{z}S_{j}^{z}
  +\sum_{\langle\langle ij\rangle\rangle}J_{2}^{ij}\mathbf{S}_{i}\cdot\mathbf{S}_{j} \nonumber \\
  &+&\frac{J_{a}}{2}\sum_{\langle ij\rangle}(\gamma_{ij}S_{i}^{+}S_{j}^{+}+\gamma_{ij}^{*}S_{i}^{-}S_{j}^{-})-\sum_{i}K_{i}(S_{i}^{z})^{2}.
\end{eqnarray}
The first and second term denote the Heisenberg exchange interactions between nearest neighbor with an Ising-type exchange anisotropy characterized by $I_{1}$.
The third term denotes the Heisenberg exchange interactions between second neighbors,
describing the interactions between spins in the same A or B sublattices (Fig.~\ref{fig1}),
characterized by $J_{2}^{A}$ and $J_{2}^{B}$, respectively.
Here we assume $J_{2}^{A}\neq J_{2}^{B}$.
The fourth term is the bond-dependent interactions~\cite{TMatsumoto} between nearest neighbor,
allowed by the symmetry and consistent with recent experiment in MnPS$_{3}$~\cite{TJHicks}.
$\gamma_{ij}=e^{i2\pi n/3}$ with $n=0,1,2$ the bond index, as illustrated in Fig.~\ref{fig1}.
The last term is the sublattice asymmetric single-ion easy-axis anisotropy,
characterized by $K_{A}$ and $K_{B}$ for the two sublattices, respectively.
The Hamiltonian above with $J_{2}^{A}=J_{2}^{B}$ and without $K_{A}$, $K_{B}$
was proposed for MnPS$_{3}$~\cite{TMatsumoto}.
The exchange anisotropy $I_{1}$, difference between $J_{2}^{A}$ and $J_{2}^{B}$,
and anisotropy field $K_{A}$, $K_{B}$ can be induced or tuned in the MnPS$_{3}$ homobilayer or in MnPS$_{3}$/CrCl$_{3}$ heterostructure,
verified in the very recent first-principles calculations~\cite{RHidalgoSacoto,FXiao}.
For negative and quite small positive $J_{2}^{A}$ and $J_{2}^{B}$, the ground state stays in collinear AFM phase~\cite{JBFouet,KHLee}.

We apply the Holstein-Primakoff transformation, $S^{z}=S-a^{\dagger}a$, $S^{+}=\sqrt{2S-a^{\dagger}a}a$,
$S^{-}=a^{\dagger}\sqrt{2S-a^{\dagger}a}$ for A-sublattice, and
$S^{z}=-S+b^{\dagger}b$, $S^{+}=b^{\dagger}\sqrt{2S-b^{\dagger}b}$,
$S^{-}=\sqrt{2S-b^{\dagger}b}b$ for B-sublattice. The Hamiltonian in Eq.~(\ref{eq1})
can be expanded as $H=\sum_{p=0}^{\infty}H_{2p}$, where $2p$ denote the number of bosonic operators.
We here keep the terms up to quartic order and neglect the ground state energy term.
With a Fourier transformation, the two-particle term can be written in the form
$H_{2}=\frac{1}{2}\sum_{\mathbf{k}}\Psi_{k}^{\dagger}H_{\mathbf{k}}\Psi_{k}$, where
$\Psi_{k}=(a_{\mathbf{k}},b_{\mathbf{k}},a_{-\mathbf{k}}^{\dagger},b_{-\mathbf{k}}^{\dagger})^{T}$, the uppercase denote the transpose.
We have
\begin{equation}\label{eq2}
  H_{\mathbf{k}}= \left(\begin{array}{cc}
  h_{\mathbf{k}} & \Delta_{\mathbf{k}} \\
  \Delta_{\mathbf{k}}^{\dagger} & h_{-\mathbf{k}}^{T}
  \end{array}\right),
\end{equation}
where $h_{\mathbf{k}}=h_{0}\mathbb{I}_{2}+\mathbf{h}\cdot\bm\sigma$,
$\Delta_{\mathbf{k}}=\bm\delta\cdot\bm\sigma$.
$\bm\sigma$ denotes the sublattice index, with
$h_{0}=(h_{A}+h_{B})/2$, $h_{x}=J_{a}S\mathrm{Re}(g_{\mathbf{k}})$,
$h_{y}=-J_{a}S\mathrm{Im}(g_{\mathbf{k}})$,
$h_{z}=(h_{A}-h_{B})/2$,
$\delta_{x}=J_{1}S\mathrm{Re}(f_{\mathbf{k}})$,
$\delta_{y}=-J_{1}S\mathrm{Im}(f_{\mathbf{k}})$,
$h_{A}=K_{A}(2S-1)+3(J_{1}+I_{1})S-J_{2}^{A}S(6-d_{\mathbf{k}})$,
$h_{B}=K_{B}(2S-1)+3(J_{1}+I_{1})S-J_{2}^{B}S(6-d_{\mathbf{k}})$,
$g_{\mathbf{k}}=1+2e^{i\frac{3}{2}k_{y}}\cos(\frac{\sqrt{3}}{2}k_{x}+\frac{2\pi}{3})$,
$f_{\mathbf{k}}=1+2e^{i\frac{3}{2}k_{y}}\cos(\frac{\sqrt{3}}{2}k_{x})$,
$d_{\mathbf{k}}=\sum_{i=1}^{6}\cos(\mathbf{k}\cdot\mathbf{a}_{i})$, $\mathbf{a}_{i}$ is the second neighbor lattice vectors.

We now discuss the effect of MMIs, i.e. the four-particle term $H_{4}$.
We have
\begin{align}\label{eq3}
  H_{4}=&-\frac{1}{N}\sum_{\{\mathbf{k}_{i}\}}\frac{J_{1}}{4}(f_{\mathbf{k}_{4}}a_{\mathbf{k_{1}}}^{\dagger}a_{\mathbf{k_{2}}}a_{\mathbf{k_{3}}}
  b_{\mathbf{k_{4}}}+f_{\mathbf{k}_{4}}^{*}b_{\mathbf{k_{1}}}^{\dagger}b_{\mathbf{k_{2}}}b_{\mathbf{k_{3}}}
  a_{\mathbf{k_{4}}} \nonumber  \\
  +&H.c.)\delta_{\{\mathbf{k}_{i}\}}^{1}+[(J_{1}+I_{1})f_{\mathbf{k}_{4}-\mathbf{k}_{2}}a_{\mathbf{k_{1}}}^{\dagger}b_{\mathbf{k_{2}}}^{\dagger}a_{\mathbf{k_{3}}}
  b_{\mathbf{k_{4}}}  \nonumber  \\
  +& \frac{J_{a}}{4}(g_{\mathbf{k}_{3}}a_{\mathbf{k_{1}}}^{\dagger}a_{\mathbf{k_{2}}}^{\dagger}b_{\mathbf{k_{3}}}
  a_{\mathbf{k_{4}}}+g_{\mathbf{k}_{1}}a_{\mathbf{k_{1}}}^{\dagger}b_{\mathbf{k_{2}}}^{\dagger}b_{\mathbf{k_{3}}}
  b_{\mathbf{k_{4}}}+H.c.)\nonumber  \\
  +&(\epsilon_{A} a_{\mathbf{k_{1}}}^{\dagger}a_{\mathbf{k_{2}}}^{\dagger}a_{\mathbf{k_{3}}}
  a_{\mathbf{k_{4}}}+\epsilon_{B}b_{\mathbf{k_{1}}}^{\dagger}b_{\mathbf{k_{2}}}^{\dagger}b_{\mathbf{k_{3}}}
  b_{\mathbf{k_{4}}}
  )]\delta_{\{\mathbf{k}_{i}\}}^{2},
\end{align}
where $\delta_{\{\mathbf{k}_{i}\}}^{1}=\delta_{\mathbf{k}_{1},\mathbf{k}_{2}+\mathbf{k}_{3}+\mathbf{k}_{4}}$,
$\delta_{\{\mathbf{k}_{i}\}}^{2}=\delta_{\mathbf{k}_{1}+\mathbf{k}_{2},\mathbf{k}_{3}+\mathbf{k}_{4}}$,
$\epsilon_{A,B}=J_{2}^{A,B}d_{\{\mathbf{k}_{i}\}}+K_{A,B}$ and $d_{\{\mathbf{k}_{i}\}}=\frac{d_{\mathbf{k}_{1}}+d_{\mathbf{k}_{4}}-2d_{\mathbf{k}_{4}-\mathbf{k}_{2}}}{4}$.
To take the many-body effect and its interplay with thermal fluctuation into account,
we employ the Green function method and define a matrix Green function as $\hat{G}(\mathbf{k},\tau)=-\langle T_{\tau}\Psi_{\mathbf{k}}(\tau)\Psi_{\mathbf{k}}^{\dagger}(0)\rangle$~\cite{ALFetter},
where $T_{\tau}$ is the chronological operator for the imaginary time $\tau$. The Heisenberg operator is defined as $A(\tau)=e^{\tau H}A(0)e^{-\tau H}$
and $H=H_{2}+H_{4}$ is Hermitian. The bracket denotes the thermal average.
To get the solution, we solve the Heisenberg equation-of-motion for the Green function elements
and apply the random phase approximation to extract the nonlinear self-energy corrections from MMIs.
After a Fourier transformation $\hat{G}(\mathbf{k},\tau)=(1/\beta)\sum_{n}\hat{G}(\mathbf{k},\omega_{n})e^{-i\omega_{n}\tau}$ with $\beta=1/T$, where $T$ is the temperature and $\omega_{n}$ is the bosonic Matsubara frequency, we can get the Dyson's equation
$\hat{G}^{-1}(\mathbf{k},\omega_{n})=i\omega_{n}\tau_{z}-H_{\mathbf{k}}^{\mathrm{eff}}$ and the effective Hamiltonian
\begin{equation}\label{eq4}
  H_{\mathbf{k}}^{\mathrm{eff}}=H_{\mathbf{k}}+\Sigma_{\mathbf{k}}=\left(\begin{array}{cc}
  h_{\mathbf{k}}^{\mathrm{eff}} & \Delta_{\mathbf{k}}^{\mathrm{eff}} \\
  (\Delta_{\mathbf{k}}^{\mathrm{eff}})^{\dagger} & (h_{-\mathbf{k}}^{\mathrm{eff}})^{T}
  \end{array}
  \right),
\end{equation}
with $h_{\mathbf{k}}^{\mathrm{eff}}=h_{0}^{\mathrm{eff}}\mathbb{I}_{2}+\mathbf{h}^{\mathrm{eff}}\cdot\bm\sigma$.
$h_{0}^{\mathrm{eff}}=(h_{A}^{\mathrm{eff}}+h_{B}^{\mathrm{eff}})/2$, $h_{x}^{\mathrm{eff}}=J_{a}\frac{(\bar{S}_{A}+\bar{S}_{B})}{2}\mathrm{Re}(g_{\mathbf{k}})$,
$h_{y}^{\mathrm{eff}}=-J_{a}\frac{(\bar{S}_{A}+\bar{S}_{B})}{2}\mathrm{Im}(g_{\mathbf{k}})$,
$h_{z}=(h_{A}^{\mathrm{eff}}-h_{B}^{\mathrm{eff}})/2$, where
$h_{A}^{\mathrm{eff}}=K_{A}(-2S-1+4\bar{S}_{A})+3(J_{1}+I_{1})\bar{S}_{B}-J_{2}^{A}\bar{S}_{A}^{\prime}(6-d_{\mathbf{k}})-h_{A}^{r}$,
$h_{B}^{\mathrm{eff}}=K_{B}(-2S-1+4\bar{S}_{B})+3(J_{1}+I_{1})\bar{S}_{A}-J_{2}^{B}\bar{S}_{B}^{\prime}(6-d_{\mathbf{k}})-h_{B}^{r}$,
$\bar{S}_{A}=S-\frac{1}{N}\sum_{\mathbf{q}}\langle a_{\mathbf{q}}^{\dagger}a_{\mathbf{q}}\rangle$,
$\bar{S}_{B}=S-\frac{1}{N}\sum_{\mathbf{q}}\langle b_{\mathbf{q}}^{\dagger}b_{\mathbf{q}}\rangle$,
$\bar{S}_{A}^{\prime}=S+\frac{1}{6N}\sum_{\mathbf{q}}d_{\mathbf{q}}\langle a_{\mathbf{q}}^{\dagger}a_{\mathbf{q}}\rangle-\frac{1}{N}\sum_{\mathbf{q}}\langle a_{\mathbf{q}}^{\dagger}a_{\mathbf{q}}\rangle$,
$\bar{S}_{B}^{\prime}=S+\frac{1}{6N}\sum_{\mathbf{q}}d_{\mathbf{q}}\langle b_{\mathbf{q}}^{\dagger}b_{\mathbf{q}}\rangle-\frac{1}{N}\sum_{\mathbf{q}}\langle b_{\mathbf{q}}^{\dagger}b_{\mathbf{q}}\rangle$,
$h_{A}^{r}=h_{B}^{r}=\frac{J_{1}}{N}\mathrm{Re}\sum_{\mathbf{q}}f_{\mathbf{q}}\langle a_{\mathbf{q}}^{\dagger}b_{-\mathbf{q}}^{\dagger}\rangle
+\frac{J_{a}}{N}\mathrm{Re}\sum_{\mathbf{q}}g_{\mathbf{q}}\langle a_{\mathbf{q}}^{\dagger}b_{\mathbf{q}}\rangle$
$(\Delta_{\mathbf{k}}^{\mathrm{eff}})_{12}=\frac{(\bar{S}_{A}+\bar{S}_{B})}{2}J_{1}f_{\mathbf{k}}-(J_{1}+I_{1})\sum_{\mathbf{q}}f_{\mathbf{k-q}}
\langle a_{\mathbf{q}}b_{-\mathbf{q}}\rangle$, $(\Delta_{\mathbf{k}}^{\mathrm{eff}})_{21}=(\Delta_{-\mathbf{k}}^{\mathrm{eff}})_{12}$.
The other terms during the random phase approximation always vanish, are thus neglected.
By diagonalizing the effective Hamiltonian in Eq. (\ref{eq4}),
$\Lambda_{\mathbf{k}}^{\dagger}H_{\mathbf{k}}^{\mathrm{eff}}\Lambda_{\mathbf{k}}=diag\{E_{\mathbf{k}},E_{-\mathbf{k}}\}$, we have
$H_{\mathbf{k}}^{\mathrm{eff}}=\sum_{\mathbf{k}}(E_{\mathbf{k}}^{\alpha}\alpha_{\mathbf{k}}^{\dagger}\alpha_{\mathbf{k}}
+E_{\mathbf{k}}^{\beta}\beta_{\mathbf{k}}^{\dagger}\beta_{\mathbf{k}})$ up to zero-point energy.
The paraunitary eigenvectors satisfy
$\Lambda_{\mathbf{k}}^{\dagger}\tau_{z}\Lambda_{\mathbf{k}}=\tau_{z}$ and $\tau_{z}$ is the Pauli matrix acting on the particle-hole space~\cite{RShindou}.
Note that the diagonalization gives us the relation $\Psi_{\mathbf{k}}=\Lambda_{\mathbf{k}}\Phi_{\mathbf{k}}$ with
$\Phi_{\mathbf{k}}=(\alpha_{\mathbf{k}},\beta_{\mathbf{k}},\alpha_{-\mathbf{k}}^{\dagger},\beta_{-\mathbf{k}}^{\dagger})^{T}$.

The additional terms in Eq.~(\ref{eq4}) compared to Eq.~(\ref{eq2}) is the self-energy term $\Sigma_{\mathbf{k}}$.
Using the relation $\Psi_{\mathbf{k}}=\Lambda_{\mathbf{k}}\Phi_{\mathbf{k}}$, the matrix element of self-energy can be expressed as
\begin{equation}\label{eq5}
  \Sigma_{\mathbf{k}}^{ij}=\frac{1}{N}\sum_{\mathbf{q},\lambda=\alpha,\beta}[T_{ij}^{\lambda}(\mathbf{k,q})n_{\mathbf{q},\lambda}(T)+Q_{ij}^{\lambda}(\mathbf{k,q})],
\end{equation}
where $n_{\mathbf{q},\lambda}$ is the Bose-Einstein distribution function $n_{\mathbf{q},\lambda}=(e^{E_{\mathbf{q}}^{\lambda}/T}-1)^{-1}$
with a zero chemical potential. The right two terms in Eq.~(\ref{eq5}) correspond to the thermal and quantum corrections, respectively.
The calculation method of them are presented in Appendix A.
Notice that the self-energy correction does not vanish even at zero temperature due to the quantum fluctuations.
Similar to the previous work~\cite{BWei,VVMkhitaryan},
the Eqs. (\ref{eq4}), (\ref{eq5}) and diagonalization relation above
form the self-consistent relations.
We can calculate the band structures at given temperature self-consistently and obtain corresponding Chern integers.
The results on temperature induced topological phases
are presented below.

\section{Results and discussions}

\subsection{Phase diagram and topological transitions}

\begin{figure}[t]
  \centering
  \includegraphics[width=0.48\textwidth]{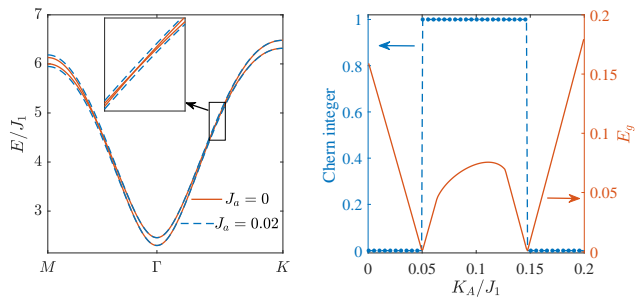}\\
  \caption{(a) The magnon band structures at $T=0$ and $K_{A}=0.1$.
  The other parameters are adopted as  $S=2.5$, $J_{1}=1.0$, $I_{1}=0.05$, $J_{2}^{A}=0.08$, $J_{2}^{B}=0.065$, $K_{B}=0.05$.
  All the parameters are in the unit $J_{1}=1.54$ meV.
  In subsequent calculations, we adopt the same parameters when not pointing out in particular. $J_{a}=0$ for solid line and $J_{a}=0.02$ for dashed lines.
  (b) The Chern integer for the acoustic branch and bandgap with respect to $K_{A}$ at $T=0$. The other parameters are the same to (a) with $J_{a}=0.02$.
  The bandgap is defined as $E_{g}=\mathrm{min}_{\mathbf{k}}(E_{\mathbf{k}}^{\beta}-E_{\mathbf{k}}^{\alpha})$ along $\Gamma-\mathbf{K}$ direction.
   }\label{fig2}
\end{figure}

We first discuss the topological phases at zero temperature $T=0$ to get an intuitive picture of  the system.
The two magnon bands are usually degenerate in pristine antiferromagnets.
The sublattice asymmetric second neighbor exchange interaction and sublattice asymmetric single-ion easy-axis anisotropy
break the band degeneracy. Especially, for $J_{2}^{A}\neq J_{2}^{B}$, the two magnon bands have
different group velocities at the same energies even when $K_{A}=K_{B}$. That is to say, the two magnon bands will separate totally or show degeneracy at limited momenta.
We find when $J_{a}=0$ and $J_{2}^{A}>J_{2}^{B}$,
in the region $0<K_{A}-K_{B}<K_{c}$, the two magnon bands will show a ring-like band intersection,
as shown in Fig.~\ref{fig2} (a) by the solid lines. Here $K_{c}\simeq 9(J_{2}^{A}-J_{2}^{B})\bar{S}^{\prime}/(4\bar{S}-2S-1)$ with
$\bar{S}=\bar{S}_{A}=\bar{S}_{B}$ and $\bar{S}^{\prime}=\bar{S}_{A}^{\prime}=\bar{S}_{B}^{\prime}$ at zero temperature.
Here we have considered the quantum corrections.
While at $K_{A}=K_{B}$ ($K_{A}=K_{B}+K_{c}$), the two bands show point touching at $\Gamma$ ($\mathbf{K}$ and $\mathbf{K}^{\prime}$) point(s) in the Brillouin zone (BZ).
Outside the above interval, the two bands are always separated and topologically trivial even for $J_{a}\neq 0$.

Finite $J_{a}$ is expected to gap the two bands and bring a nontrivial topology when $0<K_{A}-K_{B}<K_{c}$.
As $g_{\mathbf{k}}$ vanishes at $\Gamma$ and $\mathbf{K}$ points, the band point touching at $\Gamma$ ($\mathbf{K}$) point
when $K_{A}=K_{B}$ ($K_{A}=K_{B}+K_{c}$) will not be changed.
Thus the two conditions give the phase boundaries between trivial and nontrivial topological phases.
In Fig.~\ref{fig2} (a), we can see the two bands are gapped in the intersection region (dashed lines).
Fig.~\ref{fig2} (b) gives the gap evolution along the $\Gamma-\mathbf{K}$ direction.
We use the Chern integer as the topological invariant, defined as
$C_{\lambda}=\frac{1}{2\pi}\int_{BZ}B_{\lambda}^{z}d^{2}\mathbf{k}$
with Berry curvature $\mathbf{B}_{\lambda}=\nabla\times\mathbf{A}_{\lambda}$ and Berry connection
$\mathbf{A}_{\lambda}=i\mathrm{Tr}[\Gamma^{\lambda}\Lambda_{\mathbf{k}}^{\dagger}\tau_{z}(\partial_{\mathbf{k}}\Lambda_{\mathbf{k}})]$, where
$\Gamma^{\lambda}$ is the diagonal matrix taking $+1$ for $\lambda$ mode and zero otherwise.
We find the Chern integer is $1$ inside the interval $K_{B}<K_{A}<K_{B}+K_{c}$ for the acoustic branch, as presented in Fig.~\ref{fig2} (b).
When the sublattice asymmetric second neighbor exchange interaction is reversed, i.e., $J_{2}^{A}<J_{2}^{B}$, the nontrivial topological phase lies
in the interval $0<K_{B}-K_{A}<K_{c}^{\prime}$ and the Chern integer for the acoustic branch is $-1$ with
$K_{c}^{\prime}\simeq9(J_{2}^{B}-J_{2}^{A})\bar{S}^{\prime}/(4\bar{S}-2S-1)$.
In the subsequent discussions, we will focus on $J_{2}^{a}>J_{2}^{b}$ as the two cases share the same physics.

\begin{figure}[t]
  \centering
  \includegraphics[width=0.48\textwidth]{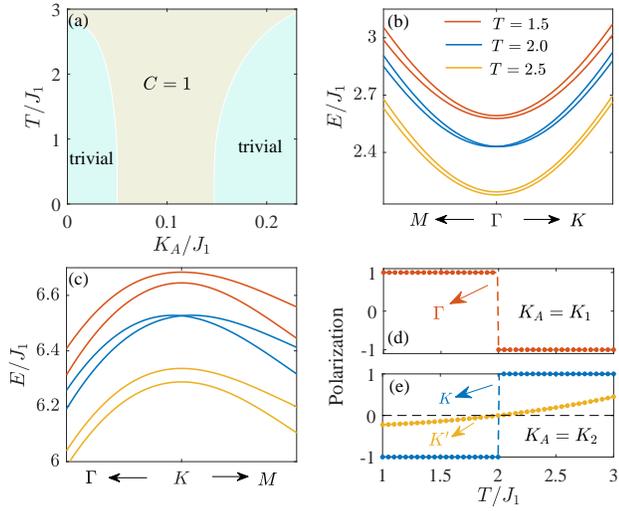}\\
  \caption{(a) The phase diagram in the $K_{A}-T$ plane. The Chern number is for the acoustic branch, while the Chern number for optical branch is opposite.
  The temperature unit is $T_{0}=J_{1}/k_{B}\simeq 17.9$ K with $J_{1}=1.54$ meV.
  (b) Magnon bands near $\Gamma$ point at three temperatures. $K_{A}=K_{1}=0.041$.
  (c) Magnon bands near $K$ point at the same three temperatures to (b). $K_{A}=K_{2}=0.169$. At $K_{1}$ and $K_{2}$,
  the critical temperature of topological transition is almost the same, $T_{c}/J_{1}\simeq 2$. (d) The magnon chirality switch at $\Gamma$ point for $K_{A}=K_{1}$.
  (e) The magnon chirality switch at $K$ point and DoP evolution at $K^{\prime}$ point for $K_{A}=K_{2}$. For all the figures, $J_{a}=0.02$.}\label{fig3}
\end{figure}

We now consider the effect of nonlinear self-energy corrections at finite temperatures and show that the trivial bands at $T=0$
will also exhibit Chern insulating phase above a critical temperature before the paramagnetic phase transitions.
We first present the phase diagram directly and choose the Chern integer of the acoustic branch as the order parameter,
calculated from the effective Hamiltonian after the self-consistent treatment at given temperatures.
The self-consistent process also helps us to confirm that the temperatures we choose are all below N\'{e}el temperature $T_{N}$.
The phase diagram in the $K_{A}-T$ plane is shown in Fig.~\ref{fig3} (a).
We can see when $K_{B}<K_{A}<K_{B}+K_{c}$, the topological phase will not be destroyed by the nonlinear corrections, same to the previous works.
Interestingly, when $K_{A}<K_{B}$ and $K_{A}>K_{B}+K_{c}$, the magnon bands show trivial phase at low temperatures
but exhibit Chern insulating phase above a critical temperature $T_{c}$, which is dependent on $K_{A}$. Note that
here $T_{c}$ is below the N\'{e}el temperature $T_{N}$.
This phenomenon is quite surprising and is not presented or discussed
in all the previous works on topological phase of magnons.
This indicates that the thermal fluctuation can indeed induce a magnonic topological insulating phase
when taking the nonlinear effect from MMIs into consideration.
Note that the phase diagram is similar to the one of topological Anderson insulators~\cite{JLi,SStutzer,EJMeier,XWLuo},
where the thermal fluctuation strength (temperature) in our system is analog to the
disorder strength.

To further investigate these thermal fluctuation induced topological transitions at finite temperature,
we plot the magnon bands at three temperatures for $K_{A}\simeq0.041 (K_{1})<K_{B}$  and $K_{A}\simeq0.169 (K_{2})>K_{B}+K_{c}$
in Fig.~\ref{fig3} (b) and (c), respectively.
The critical temperature of topological transition for the two values are almost the same,
$T_{c}/J_{1}\simeq 2$. For $K_{a}=K_{1}$, we plot the bands near $\Gamma$ point. Besides the magnon energy renormalization,
we can see the bandgap at $\Gamma$ point decreases as the temperature increases.
The spectrum become gapless at $T=T_{c}$. Further increasing temperature reopens the gap. During this process,
we have checked the two bands at other points in the BZ are always gapped.
From the phase diagram in Fig.~\ref{fig3} (a), the Chern number is $0$ below $T_{c}$ and $1$ above $T_{c}$.
The other $K_{A}<K_{B}$ value shares the same behavior with different critical temperature $T_{c}$.
Such temperature induced topological phase at the region $K_{A}<K_{B}$ is related
to the weak ferrimagnetic phase induced at finite temperatures (see Appendix B),
which arises from the imbalanced occupation number of the two magnon branches.
The zero bandgap condition at $\Gamma$ point is
$K_{A}(-2S-1+4\bar{S}_{A})+3(J_{1}+I_{1})\bar{S}_{B}=K_{B}(-2S-1+4\bar{S}_{B})+3(J_{1}+I_{1})\bar{S}_{A}$.
$\bar{S}_{A}\neq\bar{S}_{B}$ at finite temperature requires $K_{A}<K_{B}$ for the topological transition in this case.
Such temperature induced topological transition is totally due to the thermal fluctuation.
That is to say, when $K_{a}=K_{b}$, a weak perpendicular magnetic field can replace the role of easy-axis asymmetry to get
same results with a negative magnetic field. The phase diagram is presented in Appendix C.
For $K_{A}=K_{2}$, the magnon bands experience a similar behavior but the gap closes and reopens at $\mathbf{K}$ point instead, as shown in Fig.~\ref{fig3} (c).
For both cases, the thermal fluctuation induces gap closing and reopening and the
topological invariant, i.e. Chern integer jumps from $0$ to $1$.

\subsection{Magnon chirality switch}

\begin{figure}[b]
  \centering
  \includegraphics[width=0.48\textwidth]{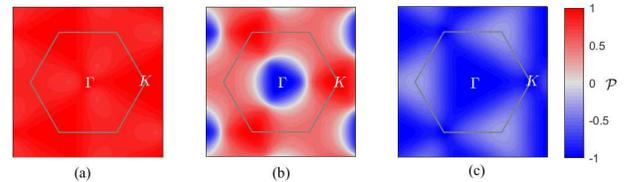}\\
  \caption{The DoP distribution in the BZ in the three regions of phase diagram Fig.~\ref{fig3} (a) for the acoustic branch.
  We adopt $K_{A}=0.02$ in (a) , $0.1$ in (b) and 0.18 in (c). The temperature is set $T=1.5$.}\label{fig4}
\end{figure}

The N\'{e}el vectors of two magnon modes precess circularly with opposite chiralities when $J_{a}=0$~\cite{SMRezende}.
For finite $J_{a}$,
the precession trajectories will become elliptical. The polarization of the magnons in antiferromagnet is similar to the one of light.
We can define the degree of polarization (DoP) for magnons in the momentum space.
At a given $\mathbf{k}$, the eigenvector at finite $J_{a}$ can be expressed as a linear combination of the polarized state at $J_{a}=0$, so we have
$\Lambda_{\mathbf{k}}^{\lambda}=\chi_{\mathbf{k},\lambda}^{+}\Lambda_{\mathbf{k}}^{+}+\chi_{\mathbf{k},\lambda}^{-}\Lambda_{\mathbf{k}}^{-}$ $(\lambda=\alpha,\beta)$, where $\Lambda_{\mathbf{k}}^{\pm}$ is the
eigenvectors for right- and left-handed precession modes when $J_{a}=0$, $\chi_{\mathbf{k},\lambda}^{\pm}$ the expansion coefficients and satisfying $|\chi_{\mathbf{k},\lambda}^{+}|^{2}+|\chi_{\mathbf{k},\lambda}^{-}|^{2}=1$.
The DoP  in momentum space is defined as $\mathcal{P}(\mathbf{k})=|\chi_{\mathbf{k},\lambda}^{+}|^{2}-|\chi_{\mathbf{k},\lambda}^{-}|^{2}$.
Below we will only focus on the acoustic branch, while the DoP for the optical branch is opposite.
At zero temperature and $J_{a}=0$,
the chirality is right-handed for $K_{A}<K_{B}$ ($\mathcal{P}(\mathbf{k})=1$) and left-handed for $K_{A}>K_{B}+K_{c}$ ($\mathcal{P}(\mathbf{k})=-1$).
Finite $J_{a}$ does not break the chirality at $\mathbf{K}$ and $\Gamma$ points. Therefore,
the chirality at $\mathbf{K}$ and $\Gamma$ points in the left trivial region in the phase diagram of Fig.~\ref{fig3} (a) is right-handed,
while at the right trivial region it is left-handed.
In the middle topologically nontrivial region,
we find the the chirality is right-handed at $\mathbf{K}$ point, while left-handed at $\Gamma$ point.
These features indicate that the topological transitions are accompanied by a magnon chirality switch at where the gap closes and reopens.
When $K_{A}<K_{B}$, the chirality
will be switched from the right-handed to left-handed across $T_{c}$, as shown in Fig.~\ref{fig3} (d).
On the other side $K_{A}>K_{B}+K_{c}$, the chirality
will be switched from the left-handed to right-handed, as shown in Fig.~\ref{fig3} (e).
Near the $\mathbf{K}^{\prime}$ point, the magnon will also experience chirality switch, same to
$\mathbf{K}$ point when $J_{a}=0$ and $K_{A}>K_{B}+K_{c}$.
For finite $J_{a}$, the evolution of the DoP at $\mathbf{K}^{\prime}$ shows a negative to positive transition [see Fig.~\ref{fig3} (e)].
As reported in the experimental works~\cite{JCenker,YNambu,YLiu,YShiota,TArakawa}, the magnon polarization in antiferromagnets
can be detected by the magneto-Raman spectroscopy~\cite{JCenker}, the polarized neutron scattering technique~\cite{YNambu}
or polarization-selective
spectroscopy~\cite{YLiu,YShiota,TArakawa}. These experimental methods greatly coincide with
our system. The detection of the magnon polarization provide experimental proofs of the temperature induced topological transitions.

We briefly discuss the DoP in the full BZ.
There are three different regions in the phase diagram in Fig.~\ref{fig3} (a).
The typical DoP distributions in the BZ for the three regions are shown in Fig.~\ref{fig4} (a) (b) (c), respectively.
In the left region, the DoP is positive [Fig.~\ref{fig4} (a)] and the right-handed mode dominates.
While in the right region, the DoP is negative [Fig.~\ref{fig4} (c)] and the left-handed mode dominates.
In the middle topological region, $\mathcal{P}$ is negative near $\Gamma$ point and positive around $\mathbf{K}$ point [Fig.~\ref{fig4} (b)].
The ring-like transition zone where $\mathcal{P}\simeq0$ is the band intersection region when $J_{a}=0$.
The differences indicate that the temperature would change the distribution,
the detection of which using the experimental techniques above provides alternative experimental proofs of our theory.

\subsection{Thermal Hall effect}

\begin{figure}
  \centering
  \includegraphics[width=0.48\textwidth]{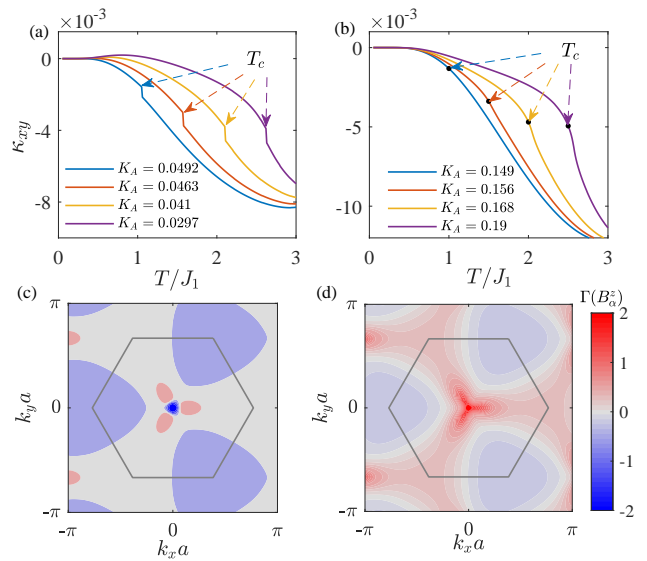}\\
  \caption{(a) (b) The thermal Hall conductivity of four different $K_{a}$ values for $K_{a}<K_{b}$ and $K_{a}>K_{c}$, respectively.
  The corresponding critical temperature $T_{c}$ is $T_{c}/J_{1}=1,1.5,2,2.5$, respectively. $\kappa_{xy}$ is in the unit $k_{B}J_{1}/\hbar=3.23\times10^{-11}$ W/K
  for $J_{1}=1.54$ meV.
  (c) (d) The distribution of Berry curvatures in log scale $\Gamma(B_{\alpha}^{z})=\textrm{sgn}(B_{\alpha}^{z})\log_{10}(1+|B_{\alpha}^{z}|)$
  for (c) $T/J_{1}=1.8$ and (d) $T/J_{1}=2.2$ at $K_{a}=0.041$. (c) (d) share the same colorbar. The gray hexagon denote the first Brillouin zone.}\label{fig5}
\end{figure}

The topological transitions from trivial to nontrivial phases indicate nontrivial transitions
of the Berry curvature distribution in the momentum space.
Therefore the topological transitions of magnons are expected to manifest themselves in the thermal transport properties~\cite{RMatsumoto,HKatsura}.
The thermal Hall conductivity is given by
\begin{equation}\label{eq6}
  \kappa_{xy}=-\frac{k_{B}^{2}T}{\hbar}\sum_{\lambda=\alpha,\beta}\int[d\mathbf{k}]B_{\lambda}^{z}(\mathbf{k})c_{2}(n_{\mathbf{k},\lambda}).
\end{equation}
where $[d\mathbf{k}]=d^{2}\mathbf{k}/(2\pi)^{2}$, $k_{B}$ is the Boltzmann constant, $\hbar$ is the reduced Planck constant,
$c_{2}(x)=(1+x)(\log\frac{1+x}{x})^{2}-(\log x)^{2}-2Li_{2}(-x)$, and $Li_{2}(x)$ is the polylogarithm function.
We plot the temperature-dependent thermal Hall conductivities $\kappa_{xy}$ in Fig~\ref{fig5} (a) and (b)
with different $K_{A}$ for comparison.
We also give the Berry curvature distribution for a typical $K_{A}$ value at
the temperature below and above $T_{c}$, respectively.

For $K_{A}<K_{B}$, we give the $\kappa_{xy}$ with four different values, corresponding to different $T_{c}$.
We can see all the $\kappa_{xy}$ show discontinuous behavior across $T_{c}$. This can be interpreted
form the Berry curvature transition near the $\Gamma$ point, as shown in Fig.~\ref{fig5} (c) and (d).
The gap closes and reopens at $\Gamma$ point and the Berry curvature experiences a jump from negative to positive values near $\Gamma$ point.
The discontinuity of  $\kappa_{xy}$ with respect to temperature is prominent and experimentally distinguishable.
Therefore the thermal Hall effect of magnons provides another route to probe the topological transition experimentally.

For $K_{A}>K_{B}+K_{c}$, we can see $\kappa_{xy}$ with respect to the temperature do not show a discontinuity near $T_{c}$ [Fig.~\ref{fig5} (b)]
This is because the thermal Hall effect of magnons are mainly contributed from the magnons with low energies.
the bandgap closes and reopen at $\mathbf{K}$ points with the energies quite large compared to $k_{B}T_{c}$.
The nontrivial Berry curvature transition near $\mathbf{K}$ point has negligible effect on $\kappa_{xy}$.
As the case $K_{A}<K_{B}$ is mostly experimentally approachable, thus the thermal Hall effect
can be the indicator of the topological transitions of magnons in realistic experiments.

\subsection{Discussions}

Above all, by challenging the belief in former works that thermal fluctuation can not induce a topological phase for magnons,
we successfully build a Chern insulating phase above finite temperatures before the paramagnetic phase transition
while the magnon bands at zero temperature is topologically trivial. The system
is the collinear antiferromagnetic insulator MnPS$_{3}$ with sublattice asymmetries induced in homobilayer or heterostructures.
We also propose realistic schemes to detect these transitions experimentally.
As the bond dependent term is equivalent to the nearest neighbor dipolar interactions,
which exist generally between the local spins in magnets.
Recent neutron resonance spin echo spectroscopy verified
the band splitting due to the dipolar interactions in MnPS$_{3}$~\cite{ARWildes,TJHicks}.
Therefore, our proposal and results should be universal in all the two-dimensional collinear antiferromagnets with sublattice asymmetries
and not limited to the honeycomb lattices. Another van der Waals magnet MnPSe$_{3}$~\cite{BLChittari,SCalder}
is also promising candidate material.
Since our proposal is based on the broken of sublattice symmetry, the ferrimagnets~\cite{MMi}, lacking it naturally,
are expected to exhibit similar temperature-induced topological phase transitions.

\section{Summary}

In summary, we demonstrate that thermal fluctuation in magnets can induce nontrivial topological phase for magnons at finite temperature
while at low temperature the bands are trivial.
The transition between trivial and nontrivial topological phases can be probed with multiple state-of-the-art techniques via measuring the thermal Hall conductivity or detecting the magnon polarization.
The temperature dependent topological phase is quite important for designing topological devices at easily achievable higher temperatures.
Our work paves the way for the study of the interplay between topological orders, MMIs and thermal fluctuation that is beyond the linear spin wave theory.

\section{Acknowledgments}

Y.-M. Li thanks Dr. Ya-Jie Wu and Dr. Bin Wei for helpful discussions.
This work is supported by the startup funding from Xiamen University.

\section*{Appendix A: the expressions for the self-energies}

We here use the Green function method and random phase approximation to get the nonlinear self-energy corrections.
The Heisenberg equation-of-motion for the Green function is given by
\begin{eqnarray*}
  \frac{d\hat{G}(\mathbf{k},\tau)}{d\tau}&=&-\delta(\tau)\tau_{z}-\langle \mathcal{T}[H,\Psi_{\mathbf{k}}(\tau)]\Psi_{\mathbf{k}}^{\dagger}(0)\rangle \\
  &=&-\delta(\tau)\tau_{z}-\tau_{z}(H_{\mathbf{k}}+\Sigma_{\mathbf{k}})\hat{G}(\mathbf{k},\tau).
\end{eqnarray*}
Here the self-energy term $\Sigma_{\mathbf{k}}$ is from the random phase approximation. With the Fourier transformation
$\hat{G}(\mathbf{k},\tau)=(1/\beta)\sum_{n}\hat{G}(\mathbf{k},\omega_{n})e^{-i\omega_{n}\tau}$, we can get
$-i\omega_{n}\hat{G}(\mathbf{k},\omega_{n})=-\tau_{z}-\tau_{z}(H_{\mathbf{k}}+\Sigma_{\mathbf{k}})\hat{G}(\mathbf{k},\omega_{n})$. Thus
\begin{equation*}
  [i\omega_{n} - \tau_{z}(H_{\mathbf{k}}+\Sigma_{\mathbf{k}})]\hat{G}(\mathbf{k},\omega_{n})=\tau_{z}.
\end{equation*}
By multiply a $\tau_{z}$ term on the both side, we can get the Dyson's equation in the main text.

From the diagonalization matrix $\Lambda_{\mathbf{k}}$, we have the
relation $a_{\mathbf{k}}=u_{\mathbf{k},a,\alpha}\alpha_{\mathbf{k}}+u_{\mathbf{k},a,\beta}\beta_{\mathbf{k}}+v_{-\mathbf{k},a,\alpha}\alpha_{-\mathbf{k}}^{\dagger}
+v_{-\mathbf{k},a,\beta}\beta_{-\mathbf{k}}^{\dagger}$, also
$b_{\mathbf{k}}=u_{\mathbf{k},b,\alpha}\alpha_{\mathbf{k}}+u_{\mathbf{k},b,\beta}\beta_{\mathbf{k}}+v_{-\mathbf{k},b,\alpha}\alpha_{-\mathbf{k}}^{\dagger}
+v_{-\mathbf{k},b,\beta}\beta_{-\mathbf{k}}^{\dagger}$. We here calculate the simplest term $\Sigma_{\mathbf{k}}^{12}$ to show how to get the relation
in Eq.~(\ref{eq5}).
As $\Sigma_{\mathbf{k}}^{12}=-\frac{J_{a}g_{\mathbf{k}}}{2}\frac{1}{N}\sum_{\mathbf{q}}(\langle a_{\mathbf{q}}^{\dagger}a_{\mathbf{q}}\rangle+\langle b_{\mathbf{q}}^{\dagger}b_{\mathbf{q}}\rangle)$. Using the relation above, we have
\begin{eqnarray*}
  \sum_{\mathbf{q}}\langle a_{\mathbf{q}}^{\dagger}a_{\mathbf{q}}\rangle &=& \frac{1}{N}\sum_{\mathbf{q}}(
  |u_{\mathbf{q},a,\alpha}|^{2}\langle\alpha_{\mathbf{q}}^{\dagger}\alpha_{\mathbf{q}}\rangle
  +|u_{\mathbf{q},a,\beta}|^{2}\langle\beta_{\mathbf{q}}^{\dagger}\beta_{\mathbf{q}}\rangle \\
  &+&|v_{-\mathbf{q},a,\alpha}|^{2}\langle\alpha_{-\mathbf{q}}\alpha_{-\mathbf{q}}^{\dagger}\rangle
  +|v_{-\mathbf{q},a,\beta}|^{2}\langle\beta_{-\mathbf{q}}\beta_{-\mathbf{q}}^{\dagger}\rangle) \\
  &=&\sum_{\mathbf{q}} [|u_{\mathbf{q},a,\alpha}|^{2} n_{\mathbf{q},\alpha}+|u_{\mathbf{q},a,\beta}|^{2}n_{\mathbf{q},\beta} \\
  &+&|v_{-\mathbf{q},a,\alpha}|^{2}(1+n_{-\mathbf{q},\alpha})+|v_{-\mathbf{q},a,\beta}|^{2}(1+n_{-\mathbf{q},\beta}] \\
  &=&\sum_{\mathbf{q},\lambda=\alpha,\beta}( |u_{\mathbf{q},a,\lambda}|^{2}+|v_{\mathbf{q},a,\lambda}|^{2})n_{\mathbf{q},\lambda}+|v_{\mathbf{q},a,\lambda}|^{2},
\end{eqnarray*}
and
\begin{equation*}
  \sum_{\mathbf{q}}\langle b_{\mathbf{q}}^{\dagger}b_{\mathbf{q}}\rangle
  =\sum_{\mathbf{q},\lambda=\alpha,\beta}( |u_{\mathbf{q},b,\lambda}|^{2}+|v_{\mathbf{q},b,\lambda}|^{2})n_{\mathbf{q},\lambda}+|v_{\mathbf{q},b,\lambda}|^{2}.
\end{equation*}
So we have
\begin{eqnarray*}
  \Sigma_{\mathbf{k}}^{12}&=&-\frac{1}{N}\sum_{\mathbf{q},\lambda=\alpha,\beta}\sum_{\xi=a,b}\frac{J_{a}g_{\mathbf{k}}}{2}[( |u_{\mathbf{q},\xi,\lambda}|^{2}+|v_{\mathbf{q},\xi,\lambda}|^{2})n_{\mathbf{q},\lambda} \\
  &+&|v_{\mathbf{q},\xi,\lambda}|^{2}]
\end{eqnarray*}
Comparing to Eq.~(\ref{eq5}), we can set $T_{12}^{\lambda}(\mathbf{k},\mathbf{q})=-\frac{J_{a}g_{\mathbf{k}}}{2}\sum_{\xi=a,b}(|u_{\mathbf{q},\xi,\lambda}|^{2}+|v_{\mathbf{q},\xi,\lambda}|^{2})$
and $Q_{12}^{\lambda}(\mathbf{k},\mathbf{q})=-\frac{J_{a}g_{\mathbf{k}}}{2}\sum_{\xi=a,b}|v_{\mathbf{q},\xi,\lambda}|^{2}$.
The other elements of the self-energy can also obtained with the same method above, then we can get the expressions in Eq.~(\ref{eq5}) in the main text.

\section*{Appendix B: temperature-induced weak ferrimagnetic phase}

\begin{figure}[t]
  \centering
  \includegraphics[width=0.48\textwidth]{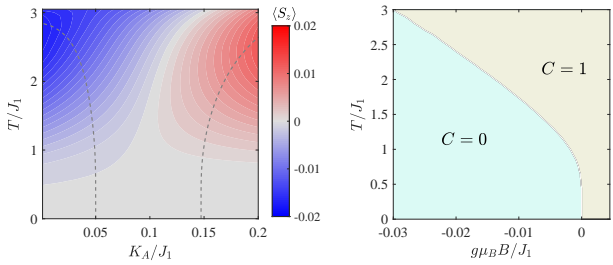}\\
  \caption{(a). The total magnetization $\langle S_{z}\rangle$ distribution in the $K_{A}$-$T$ plane.
  The dashed gray lines are the phase boundaries adopted from Fig.~3 (a).  (b).
  Phase diagram under magnetic field. $K_{A}=K_{B}=0.05$. The other parameters are the same to the main text.
  }\label{fig6}
\end{figure}

Due to sublattice asymmetries, the band degenaracies is broken.
At finite temperatures, the occupation number for the two bands is different $n_{\mathbf{q},\alpha}>n_{\mathbf{q},\beta}$.
This will induce a weak ferrimagnetic phase.
The total magnetization along $z$ direction is defined as $\langle S_{z}\rangle=\bar{S}_{A}-\bar{S}_{B}
=\frac{1}{N}\sum_{\mathbf{k}}\langle b_{\mathbf{k}}^{\dagger}b_{\mathbf{k}}\rangle-\langle a_{\mathbf{k}}^{\dagger}a_{\mathbf{k}}\rangle$.
From Fig~\ref{fig6} (a), we can see $\langle S_{z}\rangle$ does not equal to zero at relatively high temperatures.
The two band touching at $\Gamma$ point should satisfy the condition
$K_{A}(-2S-1+4\bar{S}_{A})+3(J_{1}+I_{1})\bar{S}_{B}=K_{B}(-2S-1+4\bar{S}_{B})+3(J_{1}+I_{1})\bar{S}_{A}$.
When $K_{A}=K_{B}$ and neglecting the MMIs, $\langle S_{z}\rangle=0$, the two magnon bands are always degenerate at $\Gamma$ point
although $J_{2}^{A}\neq J_{2}^{B}$.
But at finite temperature,  $\langle S_{z}\rangle \neq 0$, zero band gap condition at $\Gamma$ point satisfies for certain value of $K_{A}$ with
$K_{A}<K_{B}$, giving rise to the topological transitions at finite temperatures in the main text.

\section*{Appendix C: phase diagram under a weak magnetic field}

A weak perpendicular magnetic field can replace the role
of easy-axis anisotropy asymmetry according to the above analysis for the band gap closing condition at $\Gamma$ point in the BZ.
We set $K_{A}=K_{B}$ to verify this. The phase diagram is shown in Fig.~\ref{fig6} (b).
At zero temperature, a positive magnetic field gives a nontrivial topological phase for magnetic field, while
a negative one gives a trivial phase. Across a magnetic field dependent critical temperature, the trivial phase can also go into
nontrivial phase.

\end{document}